\documentclass[twocolumn]{aastex631}

\usepackage[utf8]{inputenc}
\usepackage{CJK}
\usepackage[utf8]{inputenc}
\usepackage[flushleft]{threeparttable}
\usepackage{longtable}
\usepackage{multirow}
\usepackage[T1]{fontenc}
\usepackage{bm}
\usepackage{array}
\usepackage{mathtools}
\makeatletter

\newcommand{\Rmnum}[1]{\expandafter\@slowromancap\romannumeral #1@}
\makeatother

\newcommand{\SolarMass}{$M_{\rm \odot}$}

\shorttitle{Massive progenitor of SN2023ixf}
\shortauthors{Fang et al.}

\begin{document}

\title{Diversity in hydrogen-rich envelope mass of type II supernovae (II): SN 2023ixf as explosion of partially-stripped intermediate massive star}
\author[0000-0002-1161-9592]{Qiliang Fang}\affiliation{National Astronomical Observatory of Japan, National Institutes of Natural Sciences, 2-21-1 Osawa, Mitaka, Tokyo 181-8588, Japan}
\author[0000-0003-1169-1954]{Takashi J. Moriya}
\affiliation{National Astronomical Observatory of Japan, National Institutes of Natural Sciences, 2-21-1 Osawa, Mitaka, Tokyo 181-8588, Japan}
\affiliation{Graduate Institute for Advanced Studies, SOKENDAI, 2-21-1 Osawa, Mitaka, Tokyo 181-8588, Japan}
\affiliation{School of Physics and Astronomy, Monash University, Clayton, Victoria 3800, Australia}
\author[0009-0000-6303-4169]{Lucía Ferrari}
\affiliation{Facultad de Ciencias Astronómicas y Geofísicas, Universidad Nacional de La Plata, Paseo del Bosque S/N, B1900FWA La Plata, Argentina}
\affiliation{Instituto de Astrofísica de La Plata (IALP), CCT-CONICET-UNLP, Paseo del Bosque S/N, B1900FWA, La Plata, Argentina}
\author[0000-0003-2611-7269]{Keiichi Maeda}\affiliation{Department of Astronomy, Kyoto University, Kitashirakawa-Oiwake-cho, Sakyo-ku, Kyoto 606-8502, Japan}
\author[0000-0003-2611-7269]{Gaston Folatelli}
\affiliation{Facultad de Ciencias Astronómicas y Geofísicas, Universidad Nacional de La Plata, Paseo del Bosque S/N, B1900FWA La Plata, Argentina}
\affiliation{Instituto de Astrofísica de La Plata (IALP), CCT-CONICET-UNLP, Paseo del Bosque S/N, B1900FWA, La Plata, Argentina}
\affiliation{Kavli Institute for the Physics and Mathematics of the Universe (WPI), The University of Tokyo, Kashiwa 277-8583, Chiba, Japan}
\author[0000-0001-7251-8368]{Keila Y. Ertini}
\affiliation{Facultad de Ciencias Astronómicas y Geofísicas, Universidad Nacional de La Plata, Paseo del Bosque S/N, B1900FWA La Plata, Argentina}
\affiliation{Instituto de Astrofísica de La Plata (IALP), CCT-CONICET-UNLP, Paseo del Bosque S/N, B1900FWA, La Plata, Argentina}
\author[0000-0002-1132-1366]{Hanindyo Kuncarayakti}
\affiliation{Tuorla Observatory, Department of Physics and Astronomy, FI-20014 University of Turku, Finland} \affiliation{Finnish Centre for Astronomy with ESO (FINCA), FI-20014 University of Turku, Finland}
\author[0000-0003-0123-0062]{Jennifer E. Andrews}
{\affiliation{Gemini Observatory/NSF's NOIRLab, 670 N. A'ohoku Place, Hilo, HI 96720, USA}
\author[0000-0002-9350-6793]{Tatsuya Matsumoto}
\affiliation{Department of Astronomy, Kyoto University, Kitashirakawa-Oiwake-cho, Sakyo-ku, Kyoto 606-8502, Japan}
\affiliation{Hakubi Center, Kyoto University, Yoshida-honmachi, Sakyo-ku, Kyoto 606-8501 Japan}

\begin{abstract}
SN 2023ixf is one of the most well-observed core-collapse supernova in recent decades, yet there is inconsistency in the inferred zero-age-main-sequence (ZAMS) mass $M_{\rm ZAMS}$ of its progenitor. Direct observations of the pre-SN red supergiant (RSG) estimate $M_{\rm ZAMS}$ spanning widely from 11 to 18\,$M_{\rm \odot}$. Additional constraints, including host environment and the pulsation of its progenitor RSG, suggest a massive progenitor with $M_{\rm ZAMS}$ > 17\,$M_{\rm \odot}$. However, the analysis of the properties of supernova, from light curve modeling to late phase spectroscopy, favor a relatively low mass scenario ($M_{\rm ZAMS}$ < 15\,$M_{\rm \odot}$). In this work, we conduct systematic analysis of SN 2023ixf, from the RSG progenitor, plateau phase light curve to late phase spectroscopy. Using \texttt{MESA}+\texttt{STELLA} to simulate the RSG progenitor and their explosions, we find that, despite the zero-age-main-sequence (ZAMS) mass of the RSG models being varied from 12.0 to 17.5\,$M_{\rm \odot}$, they can produce light curves that well match with SN 2023ixf if the hydrogen envelope mass and the explosion energy are allowed to vary. Using late phase spectroscopy as independent measurement, the oxygen emission line [O I] suggests the ZAMS is intermediate massive ($\sim$\,16.0\,{\SolarMass}), and the relatively weak H$\alpha$ emission line indicates the hydrogen envelope has been partially removed before the explosion. By incorporating the velocity structure derived from the light curve modeling into an axisymmetric model, we successfully generated [O I] line profiles that are consistent with the [O I] line observed in late phase spectroscopy of SN 2023ixf. Bringing these analyses together, we conclude that SN 2023ixf is the aspherical explosion of an intermediate massive star ($M_{\rm ZAMS}$\,=\,15\,-\,16\,$M_{\rm \odot}$) with the hydrogen envelope being partially stripped to 4\,-\,5\,{\SolarMass} prior to its explosion.
\end{abstract}
 


\section{INTRODUCTION}
SN 2023ixf is a type II supernova (SN II) discovered in the nearby galaxy M101 on May 19, 2023. After its discovery by \citet{itagaki}, SN 2023ixf attracted the attention of the community, and extensive observations were being conducted, including photometry and spectroscopy covering ultraviolet (UV), optical, to infrared (IR) bands. The explosion site is also observed by $Hubble$ Space Telescope ($HST$), $Spitzer$ Space Telescope, and ground-based telescopes. These observations confirm that the progenitor of SN 2023ixf is a dusty red supergiant (RSG), surrounded by confined circumstellar medium (CSM) (\citealt{berger23,bostroem23,chandra23,dong23,grefenstette23,hiramatsu23,hosseinzadeh23,jacobson23,kilpatrick23,mereminskiy23,panjkov23,smith23,teja23,yamanaka23,martinez24,neustadt24}). 

Being one of the most well-observed SN II, SN 2023ixf holds significant potential for testing modern theories of stellar evolution and core-collapse. For this purpose, it is important to accurately measure the zero-age-main-sequence (ZAMS) mass of its progenitor. However, the estimated $M_{\rm ZAMS}$ using different methods are inconsistent. Imaging of the progenitor prior to the explosion is one of the most direct method to estimate $M_{\rm ZAMS}$, while the estimations based on different assumptions on the properties of the dust and the models differ significantly: 11\,$\pm$\,2\,{\SolarMass} (\citealt{kilpatrick23}); 12\,-\,14\,{\SolarMass} (\citealt{vandyk24}); 16.2\,-\,17.4\,{\SolarMass} (\citealt{niu23}); 17\,$\pm$\,4\,{\SolarMass} (\citealt{jencson23}); 18.1$^{+0.7}_{-1.4}$\,{\SolarMass} (\citealt{qin23}). The monitoring of the 2023ixf progenitor with $Spitzer$ Space Telescope and ground-based telescopes reveal mid-IR variability with a period of $\sim$\,1000\,days. Making use of the period-luminosity relation of RSG in M31 (\citealt{soraisam18}), \citealt{soraisam23} estimate $M_{\rm ZAMS}$ to be 20\,$\pm$\,4\,$M_{\rm \odot}$. Further, the analysis of the stellar population in the vicinity of the explosion site favor massive progenitor, from 16.2\,$\sim$\,17.4\,{\SolarMass} (\citealt{niu23}) to around 22\,{\SolarMass} (\citealt{liu23}).

Hydrodynamic and radiative transfer modeling of the expelled material ($ejecta$) after the explosion is another useful way to constrain the properties of the progenitor. 
\citet{bersten24} shows that the plateau phase light curve can be well-fitted by the model with $M_{\rm ZAMS}$\,=\,12\,$M_{\rm \odot}$ and explosion energy $E$\,=\,1.2\,$\times$\,10$^{51}$\,erg (hereafter we refer 1.0\,$\times$\,10$^{51}$\,erg as 1.0 foe). Progenitor model with $M_{\rm ZAMS}$\,=\,15\,$M_{\rm \odot}$ cannot provide the right plateau duration and magnitude at the same time. \citet{moriya24} and \citet{singh24} employ the RSG models from \citet{kepler16},  and they also find the model with $M_{\rm ZAMS}$\,=\,10\,$M_{\rm \odot}$ and $E$\,=\,2.0\,foe best matches the light curve. The late-phase ($nebular$) spectroscopy, derived at 250 days after explosion, supports the relatively low mass scenario: when the ejecta becomes transparent, the spectroscopy is dominated by forbidden emission lines. Among them, the [O I] line can be used to measure the oxygen mass in the ejecta and constrain $M_{\rm ZAMS}$ (\citealt{fransson89,jerkstrand12,jerkstrand14,fang19,hiramatsu21,fang22}). \citet{ferrari24} found that the oxygen yield of SN 2023ixf is more consistent with $M_{\rm ZAMS}$\,=\,12\,-\,15\,$M_{\rm \odot}$.

Table 1 summarizes the inferred ZAMS mass of the progenitor of SN 2023ixf from different representative studies.

In this work, we aim to solve the inconsistency seen in pre-SN images, light curve modeling and nebular spectroscopy by bringing the uncertainty of pre-SN mass-loss into consideration. In \S 2, we construct RSG models that have the same $T_{\rm eff}$ and $L$ as pre-SN images from \citet{kilpatrick23}, \citet{vandyk24} and \citet{qin23} using \texttt{MESA}. The hydrogen-rich envelope of these RSG models are then artificially removed to mimic binary interaction or late stellar activities that may induce strong mass-loss. The partial removal of the hydrogen-rich envelope hardly change their positions on the Hertzsprung\,-\,Russell diagram (HRD) but can significantly affect the resulting light curves. The progenitor models are then used as the input of \texttt{MESA}+\texttt{STELLA} to trigger SNe explosions and the radiative transfer modeling of the light curves. In \S 3, the model light curves are compared with the observational data of SN 2023ixf, which reveals that, progenitor models with $M_{\rm ZAMS}$ larger than 15\,$M_{\rm \odot}$ can produce light curves that closely match with observation if their hydrogen-rich envelopes are partially removed to $\sim$\,4\,$M_{\rm \odot}$. Light curve modeling therefore cannot constrain $M_{\rm ZAMS}$ without knowing the amount of the hydrogen-rich envelope. In \S 4, we use nebular spectroscopy as an independent constraint on $M_{\rm ZAMS}$. By taking $\gamma$-photon leakage into consideration, we find evidence for intermediate massive star ($M_{\rm ZAMS}$\,$\sim$\,16\,$M_{\rm \odot}$) and small hydrogen-rich envelope mass ($M_{\rm Henv}\lessapprox$ 5\,$M_{\rm \odot}$). The double-peaked [O I] can be interpreted as an axisymmetric explosion. The conclusion is left to \S 5.

\begin{deluxetable}{ccc}[t]
\centering
\label{tab:literature}
\tablehead{
\colhead{Method}&$M_{\rm ZAMS}$ (\SolarMass)&References
}
\startdata
Host environment&17\,$\sim$\,19&\citet{niu23}\\
&\,$\sim$\,22&\citet{liu23}\\
\hline
&11\,$\pm$\,2&\citet{kilpatrick23}\\
&17\,$\pm$\,4&\citet{jencson23}\\
pre-SN images&16.2\,$\sim$\,17.4&\citet{niu23}\\
&18.1$^{+0.7}_{-1.4}$&\citet{qin23}\\
&12\,$\sim$\,14&\citet{vandyk24}\\
\hline
Pulsation &20\,$\pm$\,4&\cite{soraisam23}\\
&17\,$\sim$\,21&\cite{hsu24}\\
\hline
&12&\citet{bersten24}\\
Light curve&10&\citet{moriya24}\\
&10&\citet{singh24}\\
\hline
Nebular spectroscopy&$<$\,15&\citet{ferrari24}\\
\hline
\enddata
\caption{Inferred ZAMS mass of the progenitor of SN 2023ixf in the literature.}
\end{deluxetable}

\section{Numerical Setup}
We use the one-dimensional stellar evolution code, Modules for Experiments in Stellar Astrophysics ($\texttt{MESA}$, \citealt{paxton11, paxton13, paxton15, paxton18, paxton19, mesa23}), version r23.05.1, to simulate progenitor models with varied ZAMS mass, starting from pre-main-sequence phase to the moment when the mass fraction of carbon $X_{\rm C}$ in the innermost cell drops below 10$^{-3}$. In this work, our light curve analysis focuses on the plateau phase, which is not affected by late stage evolution after core carbon depletion. We employ the mixing scheme similar to \citet{martinez20}, i.e., Ledoux criterion for convection, exponential overshooting parameters $f_{\rm ov}$\,=\,0.004 and $f_{\rm ov, 0}$\,=\,0.001, semiconvection efficiency $\alpha_{\rm sc}$\,=\,0.01 (\citealt{farmer16}), thermohaline mixing coefficient $\alpha_{\rm th}$\,=\,2 (\citealt{kippenhahn80}). The mixing length parameter $\alpha_{\rm mlt}$ is varied to tune the effective temperature $T_{\rm eff}$ of the progenitors such that the RSG models match with pre-SN images on the HRD at the end point of the calculation. Throughout the calculation, we ignore wind-driven mass loss.  After core helium depletion, we turn on the command \texttt{relax\_initial\_mass\_to\_remove\_H\_env} and use \texttt{extra\_mass\_retained\_by\_remove\_H\_env} to artificially remove the hydrogen-rich envelope. The calculation is carried on until core carbon depletion without mass-loss.
The ZAMS mass range is selected to match with the luminosity from \citealt{qin23} (high mass; 16.5$\,-\,$18.5$\,M_{\rm \odot}$), \citealt{vandyk24} (intermediate mass; 14.0$\,-\,$16.0$\,M_{\rm \odot}$) and \citealt{kilpatrick23} (low mass; 11.5$\,-\,$13.0$\,M_{\rm \odot}$). Here our ZAMS mass estimation is slightly higher than \citet{vandyk24}, where the ZAMS mass is proposed to be 12.0$\,\sim\,$14.0$\,M_{\rm \odot}$. This is because compared with their reference models, given the same ZAMS mass, our models have smaller helium cores and appear to be fainter on HRD. However, we do not attempt to change $f_{\rm ov}$, a key parameter that controls the helium core mass for fixed ZAMS mass (see for example \citealt{temaj24}), to align our ZAMS mass estimation because the progenitor models in this work follow the same $M_{\rm ZAMS}$-$M_{\rm He\,core}$ relation as \citet{kepler16}, which is frequently used as the initial models for core-collapse simulation and nebular spectroscopy modeling that will be discussed in later sections. It is the helium core mass (or more precisely, the carbon-oxygen core mass), rather than the ZAMS mass that determines the amount of oxygen element and the core-collapse process.
Throughout this work, the estimation of ZAMS mass is based on these progenitor models that follow a fixed $M_{\rm ZAMS}$-$M_{\rm He\,core}$ relation (Figure~\ref{fig:HRD}). 

We further note that $T_{\rm eff}$ from \citet{vandyk24}, estimated to be 2770$\,{\rm K}$, is too cool to be reproduced by the RSG models in this work, we therefore adopt $T_{\rm eff}\,=\,{\rm3110}\,\sim\,{\rm3330}\,$K, which are the lower and upper values of $T_{\rm eff}$ of IRC-10414, a Galactic RSG analog of SN 2023ixf progenitor (\citealt{gvaramadze14,messineo19,vandyk24}).

After core carbon depletion, we closely follow the test suite \texttt{ccsn\_IIp} to trigger the explosion of the progenitor model with varied explosion energy $E_{\rm K}$. After the shock wave has reached to 0.05\,{\SolarMass} below the stellar surface, we artificially deposit 0.06\,{\SolarMass} $^{\rm 56}$Ni uniformly in the helium core ($M_{\rm Ni}$ is taken from \citealt{singh24} and \citealt{moriya24}), and use boxcar scheme to smooth the abundance profiles in the ejecta (\citealt{kasen09,dessart12,dessart13,snec15,fang24}). The model is then hand-off to \texttt{STELLA} for the calculation of the light curve. The detailed description of this workflow can be found in the literature (\citealt{paxton18,goldberg19,hiramatsu21,fang24}). 

The parameters of the RSG progenitor models are listed in Table~\ref{tab:progenitors}. The comparison of the models and the RSG progenitors of SN 2023ixf from pre-SN images on HRD are shown in Figure~\ref{fig:HRD}. 

\begin{figure}[!htb]
\epsscale{1.15}
\plotone{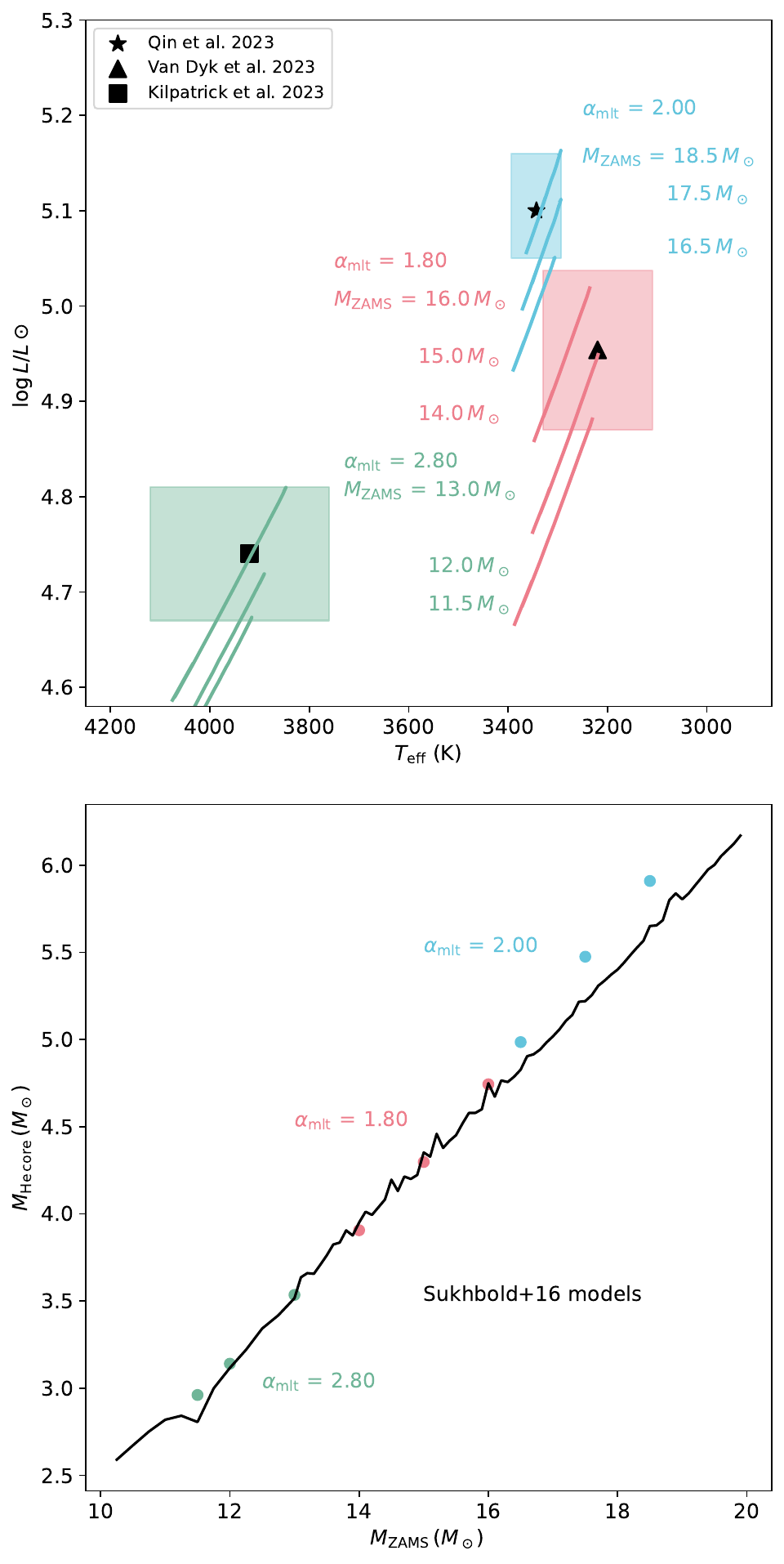}
\centering
\caption{Upper panel: The comparison of the models and the RSG progenitors of SN 2023ixf from pre-SN images on Hertzsprung\,-\,Russell diagram. The black star, triangle and square are the measurements of \citet{qin23}, \citet{vandyk24} and \citet{kilpatrick23}, which represent the high-mass, intermediate-mass and low-mass estimations for SN 2023ixf progenitor. The shaded regions are the corresponding uncertainties. The color lines are the evolution tracks of some representative progenitor models at carbon burning phase. Lower panel: The $M_{\rm ZAMS}-M_{\rm He\,core}$ relation in this work. The black line represents the models from \citet{kepler16}.}
\label{fig:HRD}
\end{figure}

\begin{deluxetable*}{cccccccc}[t]
\centering
\label{tab:progenitors}
\tablehead{
\colhead{RSG Progenitor}&$T_{\rm eff}\,$(K)&log$\,L/L_{\rm \odot}$&$\alpha_{\rm mlt}$&$M_{\rm ZAMS}\,$($M_{\rm \odot}$)&$M_{\rm Henv}\,$($M_{\rm \odot}$)&$M_{\rm rem}\,$($M_{\rm \odot}$)&$E_{\rm K}\,$(foe)
}
\startdata
\citet{qin23}&3343$_{\rm -50}^{\rm +50}$&5.10$_{\rm -0.05}^{\rm +0.05}$&2.00&17.5&3.0$\,-\,$7.5&1.8&0.5$\,-\,$1.5\\
\citet{vandyk24}&3220$_{\rm -110}^{\rm +110}$&4.95$_{\rm -0.08}^{\rm +0.09}$&1.80&15.0&3.0$\,-\,$7.5&1.8&0.5$\,-\,$1.5\\
\citet{kilpatrick23}&3920$_{\rm -160}^{\rm +200}$&4.74$_{\rm -0.07}^{\rm +0.07}$&2.80&12.0&3.0$\,-\,$8.0&1.5&1.0$\,-\,$2.5\\
\hline
\enddata
\caption{Progenitor models in this work.}
\end{deluxetable*}

\section{Light curve analysis}
The photometry data of SN 2023ixf in $BgVriz$ bands are collected from \citet{singh24}. Here, we adopt distance 6.85$\,\pm\,$0.15 Mpc (\citealt{riess22}), and total extinction $E{\rm (}B-V{\rm )}$\,=\,0.039 mag (\citealt{schlafly11,lundquist23}) with $R_{V}\,$=\,3.1. Extinctions in different bands are estimated from the extinction law of \citealt{cardelli89}.

The light curve of SN 2023ixf is characterized by a rapid rise to $M_{V} \sim -18.4$ mag, followed by a gradual decline to a plateau at $M_{V} \sim -17.6$ mag. The early phase emission indicates the presence of dense circumstellar material (CSM) that is not predicted by standard stellar evolution theory (\citealt{grefenstette23,hiramatsu23,hosseinzadeh23,jacobson23,smith23,teja23,yamanaka23,martinez24}) 
whose properties have been extensively studied (see, for example, \citealt{hiramatsu23, martinez24, singh24}). While the CSM around SN 2023ixf is crucial for understanding stellar evolution, it is not the focus of this work. As pointed out by \citet{morozova18} and \citet{moriya11}, CSM interaction dominates early-phase observations, but its effects on the plateau phase, especially in $gVriz$ bands, are minimal. The plateau duration and magnitude are mainly affected by the explosion energy ($E_{\rm K}$) and the ejecta mass ($M_{\rm eje}$). Therefore, we do not include CSM in our models to avoid introducing unrelated parameters. Our analysis is restricted to $t > 40$ days, i.e., after the midpoint of the plateau.

For progenitor models with the same $M_{\rm ZAMS}$, we use $M_{\rm Henv}$ and $E_{\rm K}$ as free parameters to fit the multi-band light curves of SN 2023ixf. The quality of the fits is evaluated from $t\,$>\,40\,days, covering roughly from the midpoint of the plateau to the onset of the radioactive tail. The ranges of $M_{\rm Henv}$ and $E_{\rm K}$ are listed in Table~\ref{tab:progenitors}, with steps of 0.25\,$M_{\rm \odot}$ 0.1\,foe, respectively. The best-fit model is determined by interpolating the model light curves to the observed epochs in the different bands and minimizing $\chi^{2}$. The photosphere velocities, estimated from the Fe II absorption minimum in early phase spectroscopy measured in \citealt{singh24}, are not included in the fitting process, but used as independent evaluations of the qualities of the fits. The results of the best-fit parameters to models with $M_{\rm ZAMS}\,=\,$17.5$\,M_{\rm \odot}$ (\citealt{qin23}), 15.0$\,M_{\rm \odot}$ (\citealt{vandyk24}) and 12.0$\,M_{\rm \odot}$ (\citealt{kilpatrick23}) are shown in Figure~\ref{fig:lc_fit}.

From the results of light curve modeling, we conclude that, if $M_{\rm Henv}$ is allowed to vary, RSG progenitors with $M_{\rm ZAMS}$ as massive as 15\,-\,18\,$M_{\rm \odot}$ can produce similar light curves to those of relatively low mass progenitors ($M_{\rm ZAMS} \sim 12~M_{\rm \odot}$) with almost all of their hydrogen-rich envelopes attached. The artificial removal of the hydrogen-rich envelope does not significantly change the stellar radius as long as the residual $M_{\rm Henv}$ remains larger than $\sim\,3\,M_{\rm \odot}$ (see Table 2 of \citealt{snec15} or Figure 1 of \citealt{fang24}). The luminosity, primarily determined by the helium core mass, is also unaffected by this process. Therefore, partial removal of the hydrogen-rich envelope does not alter the position of the progenitor RSG on the HRD, aligning with results from pre-SN images, while it introduces diversity in light curve properties. Without knowing the mass of the residual hydrogen-rich envelope, which can be significantly influenced by the presence of a companion star or complex eruptive activity in the late phases of massive star evolution, light curve modeling cannot determine $M_{\rm ZAMS}$ of SN progenitor. For this purpose, other independent measurement is required.

\begin{figure*}[!htb]
\epsscale{1.15}
\plotone{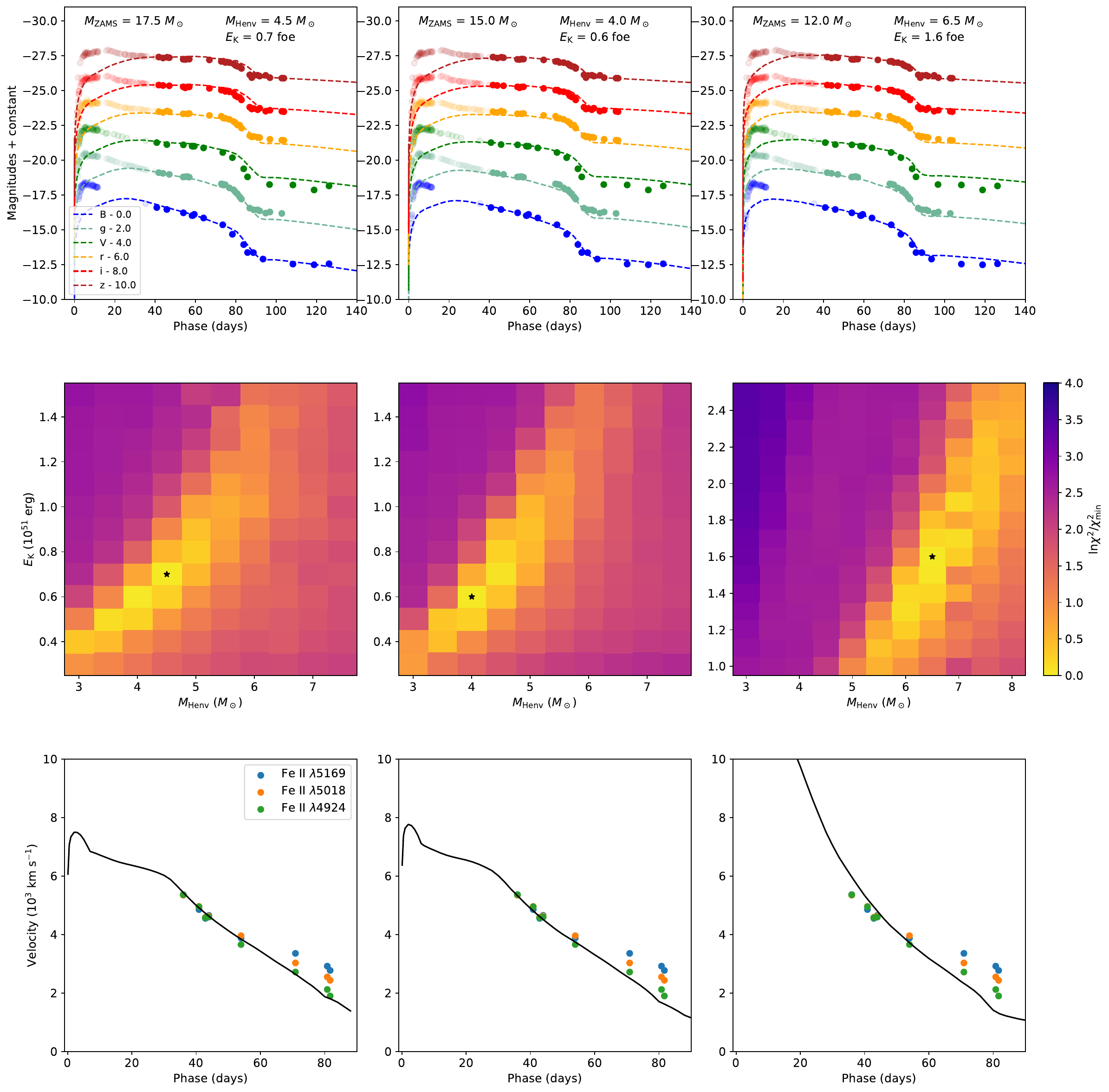}
\centering
\caption{Upper panels: The light curves of the best-fit $M_{\rm Henv}$ and $E_{\rm K}$ for models with $M_{\rm ZAMS}$ = 17.5\,$M_{\rm \odot}$ (left; \citealt{qin23}), $M_{\rm ZAMS}$ = 15.0\,$M_{\rm \odot}$ (middle; \citealt{vandyk24}) and $M_{\rm ZAMS}$ = 12.0\,$M_{\rm \odot}$ (right; \citealt{kilpatrick23}). The photometry of SN 2023ixf are the scatter points, and data from different filters are labeled by different colors. The solid points are used for fitting. The dotted lines represent the light curves of the best-fit models; Middle panels: Color-coded ratio of $\chi^{2}$ to $\chi^{2}_{\rm min}$ at $M_{\rm Henv}$-$E_{\rm K}$ space. The black star marks the parameter pair ($M_{\rm Henv}$, $E_{\rm K}$) with the minimum $\chi^{2}$. The color-bar is floored at ln\,$\chi^{2}/\chi^{2}_{\rm min}$\,=\,4.0. Lower panels: The evolution of the photospheric velocities of SN 2023ixf (scatter points) and the best-fit models (black lines).}
\label{fig:lc_fit} 
\end{figure*}

\section{Nebular Spectroscopy analysis}
While the light curve during the plateau phase is largely affected by the mass of the hydrogen-rich envelope and is therefore sensitive to the uncertain mass-loss history prior to the explosion, late-phase ($nebular$ phase) spectroscopy, taken several months to a year after the explosion when the ejecta becomes optically thin, is primarily determined by the properties of the innermost core. At this phase, the spectroscopy of the SN is dominated by emission lines, with particularly strong lines being [O I] $\lambda\lambda$6300,6363, H$\alpha$ and [Ca II] $\lambda\lambda$7291,7323. The absolute or relative flux of [O I] is an useful proxy of the amount of the oxygen elements in the core region, therefore is frequently employed as the indicator of the ZAMS mass of the progenitor from aspects of both theory and observation. In this section, we conduct analysis on the nebular spectroscopy of SN 2023ixf, taken at $t\,$=\,259 days after the explosion (\citealt{ferrari24}).
\subsection{[O I] luminosity} 
In \citet{ferrari24}, based on nebular spectroscopy analysis, the ZAMS mass of the progenitor of SN 2023ixf is proposed to be 12\,-\,15$\,M_{\rm \odot}$, consistent with the pre-SN images of \citet{kilpatrick23} and \citet{vandyk24}. The conclusion is made based on several lines of evidence: (1) when scaled to the same distance, the [O I] flux of SN 2023ixf is relatively low compared with model spectroscopy at similar phase, taken from \citealt{jerkstrand12} and \citealt{jerkstrand14}; (2) The [O I]/[Ca II] ratio, which is a useful proxy of the progenitor CO core mass, is as low as 0.51, falling between the models with $M_{\rm ZAMS}\,=\,$12\,$M_{\rm \odot}$ and $M_{\rm ZAMS}\,=\,$15\,$M_{\rm \odot}$.

While these arguments are well-supported by direct comparison with model spectroscopy, they may not fully apply to SN 2023ixf. The models employed for comparison assume massive hydrogen-rich envelopes and have been found to match well with the observations of SNe 2004et and 2012aw that have a long plateau of $\sim$ 120 days. In contrast, the plateau duration of SN 2023ixf is $\sim$ 80 days (\citealt{bersten24}), approximately 40 days shorter, meaning it enters the nebular phase earlier. Additionally, the radioactive tail of SN 2023ixf declines faster than those of SNe 2004et and 2012aw. These two factors make SN 2023ixf appear $\sim$ 0.8\,mag fainter in $R$-band than the model spectroscopy, suggesting that the low [O I] luminosity of SN 2023ixf does not necessarily indicate a low oxygen abundance compared with model progenitors. Instead, it is likely a result of a lower fraction of $\gamma$-photons, emitted from the radioactive decay of $^{56}$Co, being trapped in the ejecta.

In Figure~\ref{fig:nebular}, we compare the model spectroscopy from \citet{jerkstrand12} and \citet{jerkstrand14} with SN 2023ixf, normalizing all spectra to the integrated fluxes from 4500 to 8000\,{\rm \AA}. This normalization ensures that the models and SN 2023ixf have the same amount of deposited radioactive energy in this wavelength range. Consequently, the fractional flux of the emission lines reflects the relative abundance of the emitting elements in the line-forming region. SN 2023ixf shows apparently stronger [O I] emission than the models with $M_{\rm ZAMS}$ = 12\,$M_{\rm \odot}$ (hereafter referred to as M12 model, respectively; similarly, M15, M19 and M25 refer to the models with $M_{\rm ZAMS}$ = 15, 19 and 25\,$M_{\rm \odot}$), while its flux is between M15 and M19 models. The fractional [O I] fluxes of the models, as a function of ZAMS masses, are compared with that of SN 2023ixf in Figure~\ref{fig:nebular}. Direct interpolation gives $M_{\rm ZAMS}$\,$\sim$\,16.3\,$M_{\rm \odot}$ for SN 2023ixf, close to the upper limit of \citet{vandyk24} and the lower limit of \citet{qin23}. 

The above analysis is based on the assumption that, the $\gamma$-photon escape probability is the same throughout the ejecta, from the dense carbon oxygen core to the hydrogen-rich envelope. In this case, decreasing the total luminosity by 60\% (0.8 mag in $R$-band) will at the same time decrease the [O I] luminosity by the same amount, therefore the fractional flux of [O I] remains unchanged and can be used to determine $M_{\rm ZAMS}$. In practice, this assumption does not hold as $\gamma$-photons can more easily escape from the outermost envelope. Here we consider a limit case, i.e., all the additional leakage of $\gamma$-photons are radiated in the form of H$\alpha$ from the envelope. The oxygen fluxes of the models and SN 2023ixf are then accordingly normalized to the integrated fluxes without H$\alpha$ line, i.e., we only consider the integrated fluxes of the metal emission lines. Similar to the above analysis, interpolation gives $M_{\rm ZAMS}$\,$\sim$\,15.2\,$M_{\rm \odot}$. Our final estimation of $M_{\rm ZAMS}$ based on nebular spectroscopy therefore falls between 15.2 to 16.3\,$M_{\rm \odot}$.

While the [O I] flux of SN 2023ixf is within the range of the M12 to M25 models, its H$\alpha$ emission is noticeably weaker. Given the same total energy, the [O I] flux of SN 2023ixf is about half the value observed in the M15 and M19 models. Although the formation of H$\alpha$ is complex and influenced by many processes such as mixing, its relative weakness compared to models with a massive hydrogen-rich envelope qualitatively supports the idea that the hydrogen-rich envelope of SN 2023ixf is partially removed. The interpretation here is limited by our current lack of nebular spectroscopy models for partially stripped SNe IIP, however, a pioneering study by \citet{dessart20} shows that for SNe IIP with low-mass hydrogen-rich envelopes, the [O I] line is not significantly affected, while the fluxes of H$\alpha$ dramatically decrease and [Ca II] slightly increase (see their Fig.9.). This behavior is roughly consistent with the observations of SN 2023ixf, and explains the relatively low [O I]/[Ca II] compared with M15 model (\citealt{ferrari24}). 

In conclusion, nebular spectroscopy analysis supports the hypothesis that SN 2023ixf is the explosion of a RSG of $M_{\rm ZAMS}\sim$ 15.2 to 16.3\,$M_{\rm \odot}$, similar to the estimation of \citet{vandyk24}, with $M_{\rm Henv}\lessapprox$ 5\,$M_{\rm \odot}$ (about half the values of M15 and M19 models), which is much lower than the prediction of single stellar models evolved with standard stellar wind (\citealt{kepler16,fang24}).

\begin{figure}[!htb]
\epsscale{1.15}
\plotone{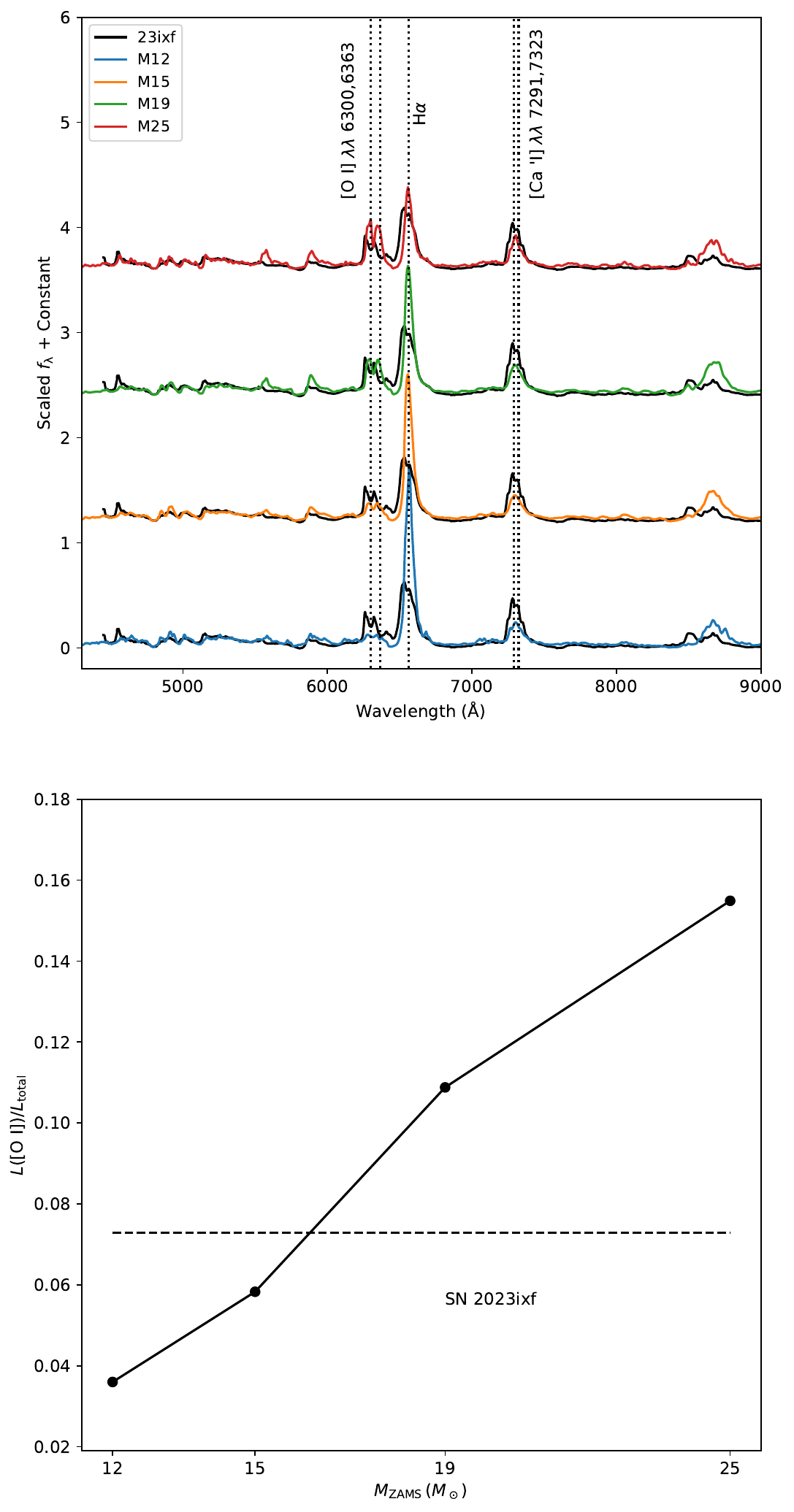}
\centering
\caption{Upper panel: The comparison of the spectroscopy models at 250 days, taken from \citet{jerkstrand12} and \citet{jerkstrand14}, with the nebular spectrum of SN 2023ixf from \citet{ferrari24}. All spectra are normalized to the integrated flux from 4500 to 8000 {\rm \AA}. Lower panel: The fractional flux of [O I] as a function of $M_{\rm ZAMS}$. The dashed line indicates the measurement for SN 2023ixf.}
\label{fig:nebular}
\end{figure}

\subsection{Emission line profiles} 
 During the nebular phase, the ejecta expands homologously, i.e., the radial expansion velocity of a fluid parcel is proportional to its radial coordinate. Additionally, the emission line widths are dominated by Doppler broadening, therefore directly reflect the spatial distributions of the emitting elements. Consequently, the emission lines observed in nebular spectroscopy provide information not only on the abundance of different elements but also on their geometric distributions within the ejecta (\citealt{taubenberger09,jerkstrand17,fang22,vanBall23}).

In this work, we focus on the profile of [O I] line. The [O I] line exhibits a horn-like (or double-peak) profile (\citealt{singh24,ferrari24}), characterized by a trough located at $v\,\sim\,$0\,km\,s$^{-1}$, with symmetric blue- and red-shifted peaks around it. While double-peaked [O I] is frequently observed in stripped-envelope supernova (SESN; core-collapse supernovae that have lost almost all of their hydrogen-rich envelope prior to the explosions; see \citealt{mazzali05,maeda08,modjaz08,taubenberger09,milisavljevic10,fang22}), it is rarely seen for the emission lines of hydrogen-rich SNe (with few exceptions; see for example \citealt{chugai05,andrews19,utrobin19,utrobin21}), although signature of asphericity can be detected from spectropolarimetry (\citealt{leonard01,leonard06,wang08,kumar16,nagao19,vasylyev23,vasylyev24,nagao24a,nagao24b}). This peculiar profile is interpreted as emission from an oxygen-rich torus surrounding a bipolar calcium-rich region, being viewed from the edge (\citealt{maeda02,maeda06,maeda08,fang22,fang24b}).

Here, we use the velocity profile from the models that best fit the light curves in \S3 to synthesize the [O I] line profile with an axisymmetric model proposed in \cite{fang24b}. The model is characterized by an oxygen-rich ball excised by two detached ellipsoids, within which all the oxygen elements are burnt into heavy elements (see Figure~\ref{fig:nebular_line_profile}; red region: explosive burning ash with $X_{\rm O}$\,=\,0; blue region: oxygen-rich unburnt material with $X_{\rm O}$\,=\,1. Here $X_{\rm O}$ is the mass fraction of oxygen). We further assume that the material in the helium core, including the oxygen-rich region, is fully mixed (\citealt{jerkstrand12}). Consequently, the boundary velocity of the oxygen-emitting region, $V_{\rm O}$, is the same as the velocity at the edge of the helium core, and the density in this region is a constant. Using the procedure outlined in \citet{fang24b}, the synthesized [O I] profiles, viewed from $\theta$ = 90 degree, are shown in Figure~\ref{fig:nebular_line_profile}. For all the models, despite variations in $M_{\rm ZAMS}$, $M_{\rm Henv}$ and $E_{\rm K}$, the synthesized [O I] profiles align well with observation. This consistency indicates that the horn-like profile of [O I] indeed originates from the oxygen-rich torus. Additionally, the [O I] line width of SN 2023ixf also requires the material in the helium core to be fully mixed: if we use the velocity at the edge of the carbon-oxygen core, which is about 1000 to 1500 km\,s$^{-1}$, to model the [O I] line, the synthesized [O I] lines are extremely narrow and none of the profiles provide a satisfactory match with the observation.

The analysis of this section is based on the assumption that the double-peak [O I] profile results from a geometrical effect of the ejecta. Nevertheless, unlike SESN where the separation of the two peaks is wide (see \citealt{fang24b}), for SN 2023ixf, the separation is $\sim$ 3000\,km\,s$^{-1}$, i.e., close to that of the two components of the [O I] doublet. We therefore cannot reject other possibilities such as clumping (\citealt{kuncarayakti20,ferrari24}) or an unipolar oxygen-rich blob moving toward the observer (\citealt{taubenberger09,milisavljevic10}), accelerated by neutron star kick (see for example \citealt{burrows24}). However, these explanations would require some SNe IIP exhibit horn-like [O I] profile with both peaks red-shifted, which, to our knowledge, have not been observed yet. In this work, we consider bipolar explosion with a torus-like structure as more plausible interpretation, but emphasize that other scenarios, such as clumping or unipolar blob, can not be exclusively rejected.

\begin{figure*}[!htb]
\epsscale{1.15}
\plotone{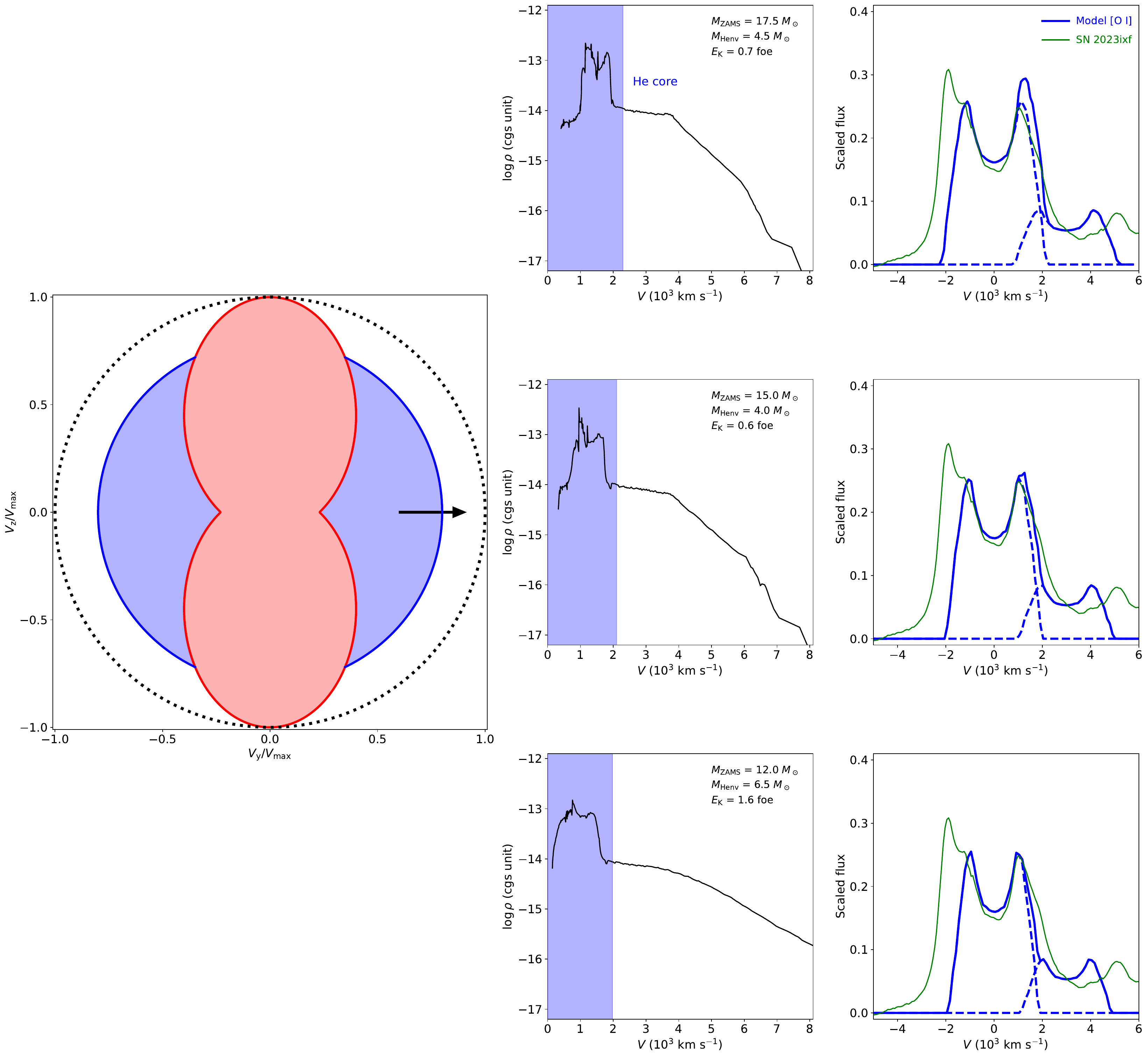}
\centering
\caption{The bipolar structure of the ejecta and the synthesized [O I] line. Left panel: The axisymmetric bipolar ejecta model proposed in \citet{fang24b}. The red shaded region is the explosive burning ash, containing only iron-peak elements. The blue shaded region is the oxygen-rich material from which [O I] is emitted; Middle panels: The density structure of the ejecta of the best-fit models for light curve modeling (\S 3 and Figure~\ref{fig:lc_fit}), as function of the velocity coordinate. The blue shaded regions represent the helium cores; Right panels: Comparison between the observed [O I] profile of SN 2023ixf (green line) with the synthesized [O I] profile (blue line). The blue dashed lines are the two components of [O I] centering at 6300 and 6363\,{\rm \AA} respectively. The ratio of their intensities are assumed to be 3:1. From top to bottom: $M_{\rm ZAMS}$ = 17.5\,$M_{\rm \odot}$ (\citealt{qin23}), $M_{\rm ZAMS}$ = 15.0\,$M_{\rm \odot}$ (\citealt{vandyk24}) and $M_{\rm ZAMS}$ = 12.0\,$M_{\rm \odot}$ (\citealt{kilpatrick23}).}
\label{fig:nebular_line_profile}
\end{figure*}

\section{Conclusion}
In this work, we construct RSG models that occupy the same position on the HRD as the proposed progenitor of SN 2023ixf, based on the pre-explosion images from \citet{kilpatrick23}, \citet{vandyk24}, and \citet{qin23}. From these progenitor models, we artificially remove their hydrogen-rich envelope and trigger explosions, and compare the resulting light curves with the multi-band photometry of SN 2023ixf. Our findings indicate that, by varying the hydrogen envelope mass $M_{\rm Henv}$ and explosion energy $E_{\rm K}$, RSG models with $M_{\rm ZAMS}$ ranging from 12.5 to 17.5\,$M_{\rm \odot}$ can produce light curves that closely match the observed data. Consequently, light curve modeling alone cannot effectively constrain $M_{\rm ZAMS}$ due to this degeneracy.

To address this limitation, we employ nebular spectroscopy as an independent method for estimating $M_{\rm ZAMS}$. The fractional flux of the [O I] line suggests $M_{\rm ZAMS}$ values between 15.2 and 16.3\,$M_{\rm \odot}$. Interestingly, the H$\alpha$ line also provides additional constraints: RSG model with $M_{\rm ZAMS}$\,=\,12.0\,$M_{\rm \odot}$ must retain a massive hydrogen envelope ($M_{\rm Henv}$\,=\,6.5\,$M_{\rm \odot}$) to match the plateau light curve. However, this is inconsistent with the weak H$\alpha$ line observed in the nebular phase, implying that a large fraction of the hydrogen-rich envelope was removed prior to the explosion as suggested by the light curve modeling results for RSG models with $M_{\rm ZAMS}$\,=\,15.0 and 17.5\,$M_{\rm \odot}$.

Finally, we employed the axisymmetric ejecta structure from \citet{fang24b} to model the [O I] line profile of SN 2023ixf. By assuming the maximum velocity of the [O I] emitting region corresponds to the edge velocity of the helium core, taken from the velocity profiles of models that best fit the observed plateau light curve, we achieve satisfactory matches between the observed double-peaked [O I] of SN 2023ixf and the synthesized [O I] profiles, viewed from 90 degrees. This agreement not only confirms the aspherical nature of the explosion, but also provides additional constraints on material mixing: the helium core material, including oxygen-rich regions, must be thoroughly mixed to account for the relatively broad [O I] profile observed in the nebular phase.

Bringing these lines of evidence together, we propose that SN 2023ixf represents the aspherical explosion of a partially stripped, intermediate-mass RSG with $M_{\rm ZAMS}$ between 15.3 and 16.2\,$M_{\rm \odot}$. We further note that stars within this mass range do not have strong stellar winds necessary to strip its hydrogen-rich envelope to this small amount. Other mechanisms, such as binary interaction (see for example \citealt{ercolino23} for a recent study), pulsation-driven mass-loss (see \citealt{yoon10_pulse}), among other potential candidates, must be involved to assist the removal of a significant fraction of the hydrogen-rich envelope.      

During the drafting of this manuscript, \citet{hsu24} presented their analysis on SN 2023ixf. Using the RSG model grid from \citet{hiramatsu21}, they found that, by varying the explosion energy, models with $M_{\rm Henv}\,\sim\,$3\,$M_{\rm \odot}$ can produce light curves that well match with observation, despite $M_{\rm ZAMS}$ varies from 15 to 22.5\,$M_{\rm \odot}$. They further use the pre-SN variability of the progenitor to constrain the properties of the progenitor, and conclude that, RSG models with $M_{\rm ZAMS}\>$>\,17\,$M_{\rm \odot}$, $R$\,>\,950\,$R_{\rm \odot}$ and $M_{\rm Henv}$\,<\,3\,$M_{\rm Henv}$ can explain the pulsation period ($\sim$\,1100 days) as well as reproduce the observed multiband light curve of SN 2023ixf. From their Table 1., the favored models have helium core mass $M_{\rm He\,core}$ > 5.5\,$M_{\rm \odot}$, or $M_{\rm ZAMS}$\,>\,18.3\,$M_{\rm \odot}$ using the $M_{\rm ZAMS}$-$M_{\rm He\,core}$ relation of \texttt{Kepler} RSG models (\citealt{kepler16}). This value is not favored by the nebular spectroscopy analysis presented in \citet{ferrari24} and this work. However, the [O I] flux is mainly determined by the oxygen mass in the ejecta. Given the same $M_{\rm He \,core}$, \texttt{MESA} seems to predict systematically lower carbon-oxygen core mass ($M_{\rm CO\,core}$) than \texttt{Kepler} (see the comparison between \citealt{temaj24} and \citealt{kepler16}), possibly due to different treatments on the microphysics of the two codes. Using the $M_{\rm He\,core}$-$M_{\rm CO\,core}$ presented in \cite{temaj24}, the models from \citet{hsu24} with $M_{\rm ZAMS}$\,=\,17.5 to 18.0\,$M_{\rm \odot}$ will have $M_{\rm CO\,core}$\,=\,3.65 to 3.90\,$M_{\rm \odot}$, translating into $M_{\rm ZAMS}$\,=\,16.6 to 17.3\,$M_{\rm \odot}$ for \texttt{Kepler} models. Given the uncertainties in the nebular spectroscopy model and direct interpolation, we consider this $M_{\rm ZAMS}$ range matches with our estimation. However, the other two models (20.5M\_eta1.5\_alpha1.5 and 21.5M\_eta1.5\_alpha1.5) can be ruled out. Combining with nebular spectroscopy analysis, we narrow down the $M_{\rm ZAMS}$ range from \citet{hsu24} to 17.5 to 18.0\,$M_{\rm \odot}$. In conclusion, the $M_{\rm ZAMS}$ of SN 2023ixf progenitor should be around 15.0 to 18.0\,$M_{\rm \odot}$.

\software{$\texttt{MESA}$ \citep{paxton11, paxton13, paxton15, paxton18, paxton19}; SciPy \citep{scipy}; NumPy \citep{numpy}; Astropy \citep{astropy13,astropy18}; Matplotlib \citep{matplotlib}}


{}
\end{document}